\documentclass[12pt,letterpaper]{article}
\usepackage{amssymb}
\usepackage{amsmath}
\usepackage{amscd}
\usepackage{graphicx}
\usepackage{tikz}
\usepackage{subfigure}
\usepackage{array}
\usepackage{pdflscape}

\def\be{\begin{equation}}
\def\ee{\end{equation}}
\def\bea{\begin{eqnarray}}
\def\eea{\end{eqnarray}}
\def\bse{\begin{subequations}}
\def\ese{\end{subequations}}

\def\vCent#1{\vcenter{\hbox{\hss#1\hss}}}

\def\bna{\bql\begin{array}{rcl}}
\def\ena{\end{array}\eql}
\def\bnn{\beq\begin{array}{rcl}}
\def\enn{\end{array}\ee}
\def\bet{\begin{tabular}}
\def\bsf{\sffamily\bfseries}

\def\4{\text{\bsf-}}

\def\fc#1#2{\relax\ifmmode{\scriptstyle\frac{#1}{#2}} 
                    \else$\scriptstyle\frac{#1}{#2}$\fi}    

\definecolor{Red}    {rgb}{1.00,0.00,0.00} 
\definecolor{Green}  {rgb}{0.00,0.75,0.00} 
\definecolor{Blue}   {rgb}{0.00,0.00,1.00} 
\definecolor{Orange} {rgb}{1.00,0.67,0.00} 
\definecolor{Purple} {rgb}{0.50,0.00,0.50} 
\definecolor{Gold}   {rgb}{1.00,0.90,0.00} 
\definecolor{Magenta}{rgb}{1.00,0.00,1.00} 
\definecolor{Turque} {rgb}{0.00,0.90,0.90} 
\definecolor{Seaweed}{rgb}{0.00,0.25,0.00} 
\definecolor{Brown}  {rgb}{0.50,0.13,0.00} 
\definecolor{Cobalt} {rgb}{0.00,0.00,0.50} 
\definecolor{Sage}   {rgb}{0.00,0.50,0.38} 
\definecolor{grey1}  {rgb}{0.20,0.20,0.20} 
\definecolor{grey2}  {rgb}{0.40,0.40,0.40} 
\definecolor{grey3}  {rgb}{0.60,0.60,0.60} 
\definecolor{grey4}  {rgb}{0.80,0.80,0.80} 
\definecolor{grey5}  {rgb}{0.90,0.90,0.90} 
\def\C#1#2{{\ifcase#1\or
             \color{Red}\or\color{Green}\or\color{Blue}\or
              \color{Orange}\or\color{Purple}\or\color{Gold}\or
             \color{Magenta}\or\color{Turque}\or\color{Seaweed}\or
               \color{Brown}\or\color{Cobalt}\or\color{Sage}\or
                 \color{grey1}\or\color{grey2}\or\color{grey3}\or
                 \color{grey4}\else\color{grey5}\fi#2}}
\definecolor{gray}{rgb}{.7,.7,.7}
\def\XXX{\colorbox{yellow}{\color{red}\bf X\kern-4pt{\Large$\bs*$}\kern-4.125ptX}}
\def\b{{\beta}}

\def\[{\left[}
\def\]{\right]}

\font\ro=cmsy10                          
\def\kcr{{\hbox{\ro \char'170}}}                
\def\ktl{{\hbox{\ro \char'170}}}        
\def\ktr{{\hbox{\ro \char'170}}}        
\def\kbl{{\hbox{\ro \char'170}}}        
\def\kbr{{\hbox{\ro \char'170}}}        

\parskip=\medskipamount                 
\thicklines                         

\newskip\humongous \humongous=0pt plus 1000pt minus 1000pt
\def\caja{\mathsurround=0pt}
\def\eqalign#1{\,\vcenter{\openup2\jot \caja
        \ialign{\strut \hfil$\displaystyle{##}$&$
        \displaystyle{{}##}$\hfil\crcr#1\crcr}}\,}
\newif\ifdtup

\def\border{                                            
        \setlength{\unitlength}{1mm}
        \newcount\xco
        \newcount\yco
        \xco=-21
        \yco=12
        \begin{picture}(140,0)
        \put(\xco,\yco){$\ktl$}
        \advance\yco by-1
        {\loop
        \put(\xco,\yco){$\kcr$}
        \advance\yco by-2
        \ifnum\yco>-240
        \repeat
        \put(\xco,\yco){$\kbl$}}
        \xco=158
        \yco=12
        \put(\xco,\yco){$\ktr$}
        \advance\yco by-1
        {\loop
        \put(\xco,\yco){$\kcr$}
        \advance\yco by-2
        \ifnum\yco>-240
        \repeat
        \put(\xco,\yco){$\kbr$}}
               \put(-19.75,13){\tiny **University of Maryland * Center for String and
         Particle  Theory * Physics Department**University of Maryland * Center
        for String and Particle  Theory** }
        \put(-19.75,-241.5){\tiny **University of Maryland * Center for String and
         Particle  Theory * Physics Department**University of Maryland * Center
        for String and Particle  Theory** }
        \end{picture}
        \par\vskip-8mm}

\def\headpic{                                           
        \indent
        \setlength{\unitlength}{.4mm}
        \thinlines
        \par
        \begin{picture}(29,16)
        \put(165,16){\line(1,0){4}}
        \put(170,16){\line(1,0){4}}
        \put(180,16){\line(1,0){4}}
        \put(175,0){\line(1,0){4}}
        \put(180,0){\line(1,0){4}}
        \put(185,0){\line(1,0){4}}
        \put(169,0){\line(0,1){16}}
        \put(170,0){\line(0,1){16}}
        \put(179,0){\line(0,1){16}}
        \put(180,0){\line(0,1){16}}
        \put(184,0){\line(0,1){16}}
        \put(185,0){\line(0,1){16}}
        \put(169,16){\oval(8,32)[bl]}
        \put(170,16){\oval(8,32)[br]}
        \put(179,0){\oval(8,32)[tl]}
        \put(185,0){\oval(8,32)[tr]}
        \end{picture}
        \par\vskip-6.5mm
        \thicklines}
\def\endtitle{\end{quotation}\newpage}                  

\begin{document}

\border\headpic {\hbox to\hsize{\today \hfill
{PP 012-015}}}
\par \noindent
\par

\par

\setlength{\oddsidemargin}{0.5in}
\setlength{\evensidemargin}{-0.5in}
\begin{center}
\vglue .10in
{\large\bf Adinkra Isomorphisms and `Seeing' Shapes with Eigenvalues}
\\[.5in]
Keith\, Burghardt\footnote{keith@umd.edu} and S.\, James Gates, Jr.\footnote{gatess@wam.umd.edu}
\\[1.2in]

{\it Center for String and Particle Theory\\
Department of Physics, University of Maryland\\
College Park, MD 20742-4111 USA}\\[1.9in]

{\bf ABSTRACT
}
\\[.01in]
\end{center}
\begin{quotation}
{
We create an algorithm to determine whether any two graphical representations 
(adinkras) of equations possessing the property of supersymmetry in one 
or two dimensions are isomorphic in shape.  The algorithm is based on the 
determinant of `permutation matrices' that are defined in this work and derivable for 
any adinkra.
}

${~~~}$ \newline ${~~~}$ \newline
\endtitle


\section{Introduction}

$~~~$ One or two dimensional spacetimes, complete set of supersymetrical (SUSY) equations 
can always be represented by graphs called adinkras \cite{Adnk1}. As Fig. \# 1 shows, viewing
these graphs can easily determine whether the two corresponding sets of equations are isomorphic.
For example, the two equation sets represented in Fig. \# 1 
\begin{figure}[!htbp]
\centering
\subfigure[]{\label{f:diamondPath}
\includegraphics[width=0.4\columnwidth]{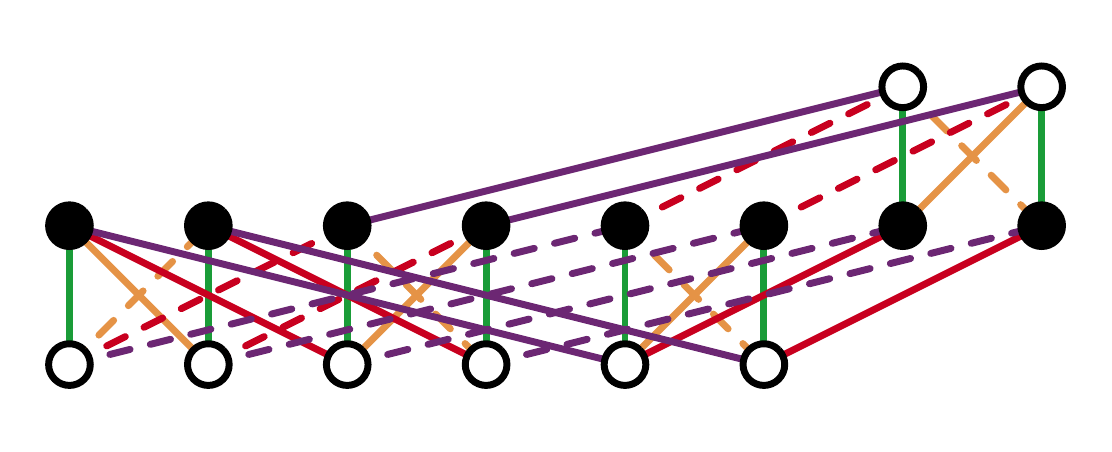}}
\quad
\subfigure[]{\label{f:bowtiePath}
\includegraphics[width=0.4\columnwidth]{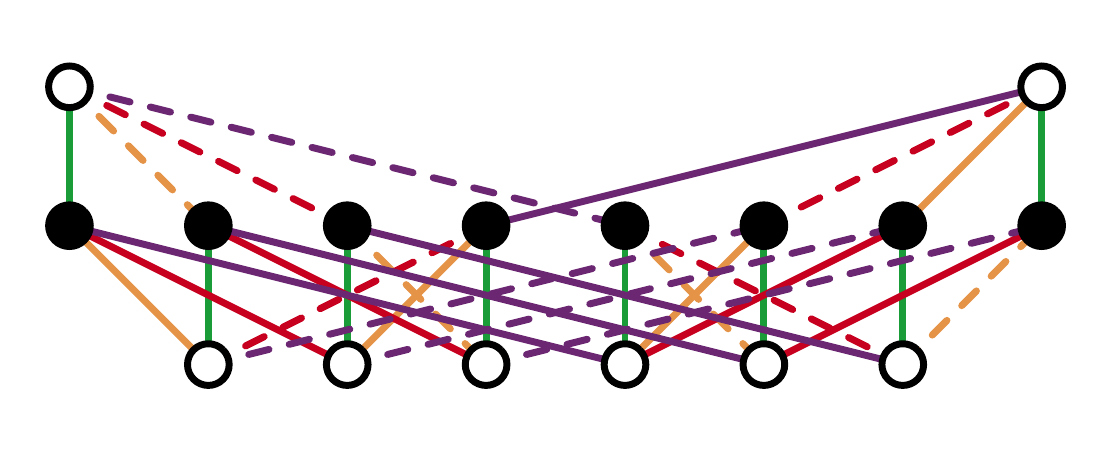}} 
\label{f:EquivAdnkEx}
\caption{Two four-color inequivalent (6$|$8$|$2) adinkras.}
\end{figure} \vskip.01in \noindent 
are clearly not isomorphic. 
One would need to look at sixty-four separate equations to prove this by conventional means.  
This example leads to the question of how to determine whether two arbitrary adinkra graphs 
are isomorphic  in shape to each other\footnote{From here on,  we will describe an algorithm that 
determines whether two adinkras are isomorphic \newline $~~~$ $~~$ as an `isomorphism 
algorithm.'  Since we are not constructing new adinrka isomorphisms, there 
\newline $~~~$ $~~$  
 should be no ambiguity in this label.}, especially when the graphs themselves may be too 
 complicated to visually inspect. In this paper, we will demonstrate an algorithm which can 
 easily determine shape isomorphisms in a computer friendly manner.

From past investigations, we know simple ways of determining adinkra 
isomorphisms become increasingly unwieldy for adinkras of increasing 
complicated structure (see Ref. \ \cite{CounterExample} for such an example), 
therefore previous work \cite{DGW} created an algorithm which could determine
isomorphisms for any adinkra. The algorithm is computationally inefficient
due to its unnecessarily complicated structure. Therefore, the 
 need for dependable and efficient algorithms that are computationally 
 simple is well motivated.

The new algorithm presented in this work for determining adinkra shape isomorphisms 
is a generalization of a previously presented algorithm \cite{AdnkEng} used to consistently
describe a set of supersymmetrical equation describing a two dimensional system.  We 
introduced in this work `permutation matrices' which, when multiplied by a super vector of bosons 
and fermions, $\mathbf{\Phi} \oplus \mathbf{\Psi}$, re-creates the supersymmetrical equations (where 
the elements in $\mathbf \Phi$ and $\mathbf \Psi$ may be each arbitrarily ordered separately). 

We then multiply these permutation matrices together to create a `total permutation matrix.' 
By calculating the eigenvalues of this matrix for any given adinkra and comparing its 
eigenvalues with the eigenvalues of any other adinkra's total permutation matrix it turns out
to be sufficient to determine if two adinkras possess the same shape.  This is analogous 
lining up the teeth of two keys to determine if they belong to the same lock.  If we set all 
 the variables in the permutation matrices to 1 and take the trace, we re-create the
``chromocharacters" \cite{G-1}, which can determine whether, by node redefinitions,
 two adinkras can have the same line dashing \cite{DGW}. These two properties allow one to
 determine whether two adinkras are isomorphic.
 
Our paper is organized as follows. In section 2, we will first introduce the new algorithm, and prove 
it will always uniquely determine adinkras. In section 3, we will briefly review the previous way 
of determining if adinkras are isomorphic, while in section 4, we determine the efficacy of this
 algorithm over the previous one. Lastly, in section 5, we demonstrate the power of the new
algorithm, by comparing two sets of adinkras using both the new and previous algorithm. 

\vskip0.7in

\section{The Isomorphism Algorithm}

$~~~$
We first describe the isomorphism algorithm in the cases of the simplest 
adinkras (see Ref.\ \cite{Adnk1,G-1} for definitions of adinkra graphs), and then 
show why it can uniquely describe adinkras. \newline \indent
The simplest inequivalent two-color adinkras appear in Fig. \# 2. 
\begin{figure}[!htbp]
\centering
\subfigure[]{\label{f:bowtie}
\includegraphics[width=0.3\columnwidth]{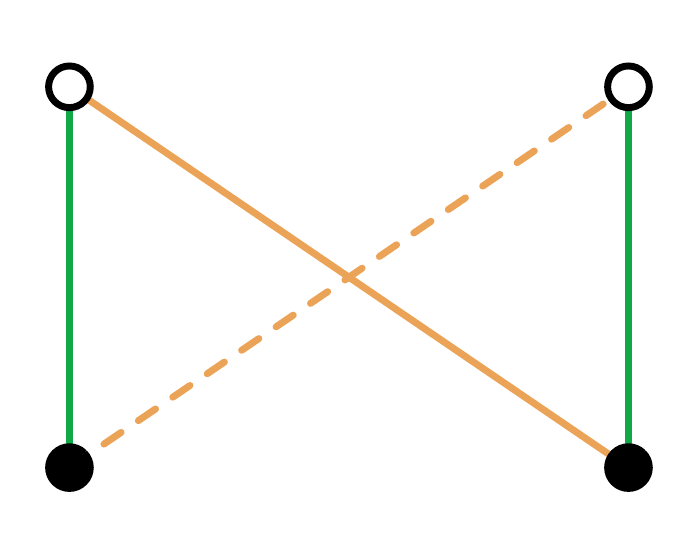}}
\quad
\subfigure[]{\label{f:diamond}
\includegraphics[width=0.3\columnwidth]{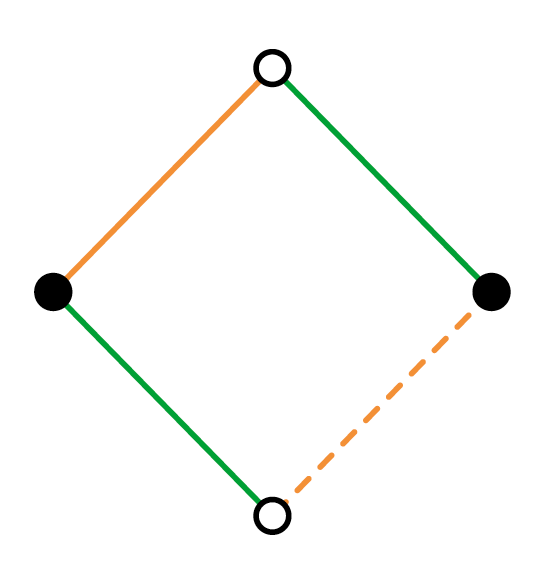}} 
\label{f:diambt}
\caption{The bow tie (left) and the diamond  (right) two-color closed paths.}
\end{figure}
\noindent
We use these to create the monochromatic `color matrices,' before we multiply them 
together, to create the `total permutation matrix' necessary for the algorithm.  The 
diamond adinkra is obtained from the bow tie adinkra by `raising' the bosonic 
2-node on the lower right of the bow tie adinkra.
The bow tie adinkra (to the left) is also an example of a `valise' adinkra.  By definition,
these are adinkras where all the bosonic nodes appear at the same height and all the
fermionic nodes appear at the same height but where the fermionic height is different 
from the bosonic height.  This naturally leads to a numbering of the nodes lexicographically 
as shown. 

First, to determine adinkra shape, we will ignore line dashing. Line dashing,
under well recognized conditions \cite{G-1}, describes the difference between 
a supermultiplet and its twisted version. This is the reason why we will later on record 
the adinkra's chromocharacters, which differentiate line dashing isometries.
 A point to note is that the matrices associated with these adinkras
have the property of possessing a single factor in each row and each
column and are thus monomials in the mathematical sense.  In addition, these matrices
are elements of the permutation group. Upon eliminating the line dashing, the two 
adinkra graphs above take the forms below in Fig. \# 3.
\begin{figure}[!htbp]
\centering
\subfigure[]{\label{f:diamondPath}
\includegraphics[width=0.3\columnwidth]{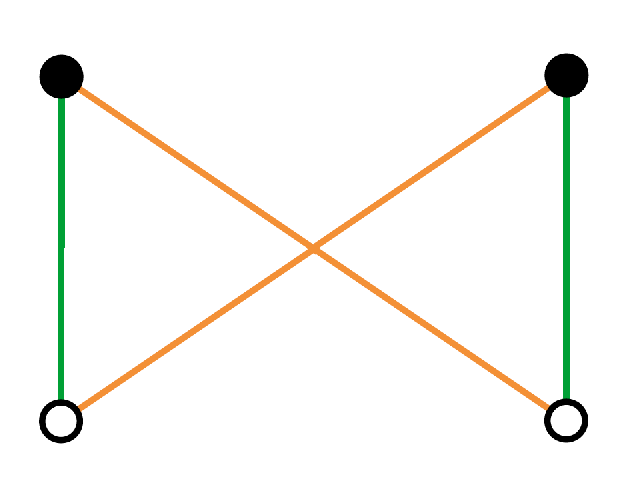}}
\put(-97,1){$1$}      \put(-24,1){$2$}
\put(-97,86){$1$}      \put(-24,86){$2$}
\quad
\subfigure[]{\label{f:bowtiePath}
\includegraphics[width=0.3\columnwidth]{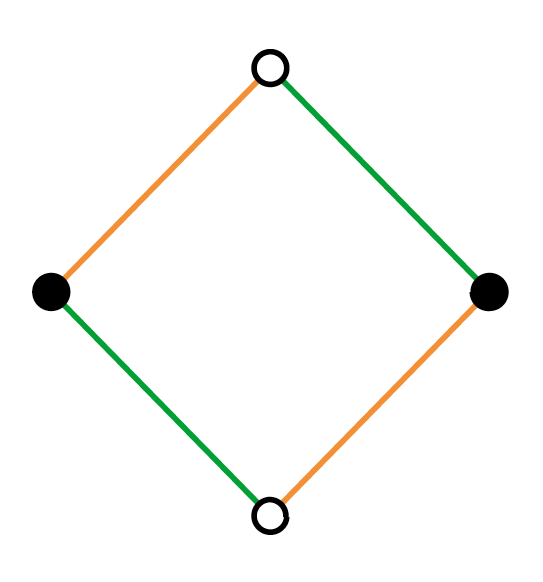}} 
\put(-51,3){$1$}         \put(-51,112){$2$}
\put(-93,59){$1$}      \put(-2,59){$2$}
\label{f:diambtPath}
\caption{Bow tie and diamond two-color closed paths with line dashing removed.}
\end{figure}
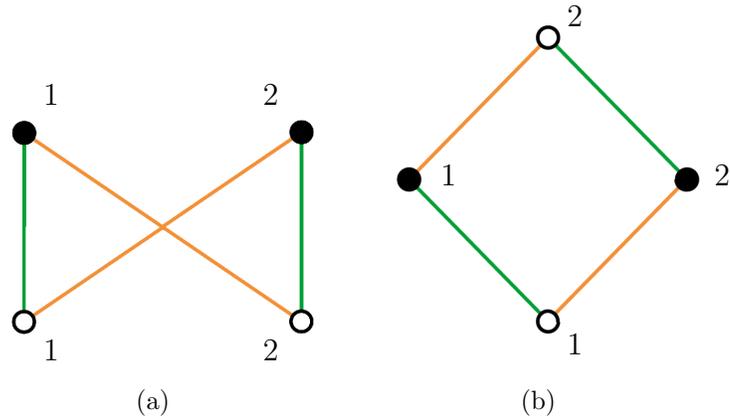
To form color matrices (each of which is dependent on line color), we first decompose 
an adinkra into its monochromatic components, as shown in Fig. \# 4 for the bow tie, 
and in Fig. \# 5 for the diamond.

\begin{figure}[!htbp]
\centering
\includegraphics[width=0.8\columnwidth]{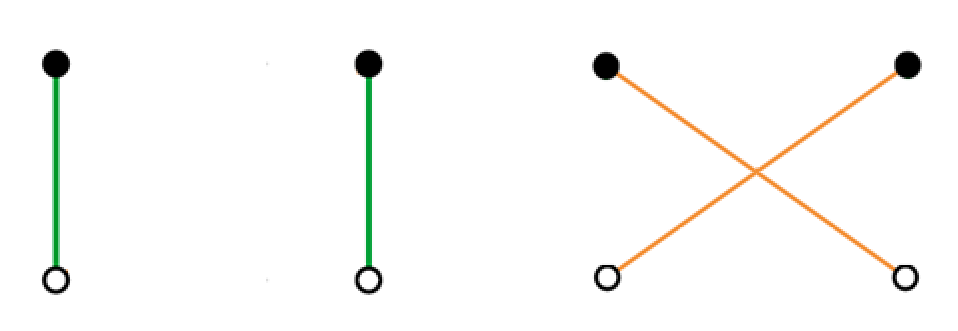} 
\label{f:bowtieN}
\caption{The monochromatic bow tie edges.}
\end{figure}
In these diagrams, we introduce $N$ distinct parameters $\b{}_{{}_{\rm I}}$, one for 
each color.  We assign a value of $\b{}_{{}_{\rm I}}$ or $\b{}_{{}_{\rm I}}^{-1}$ depending  
on whether the colored line segment is located above or below the open node attached 
to it. The $\b{}_{{}_{\rm I}}^{\pm1}$ assignment is equivalent to multiplying a fermion or 
boson field, by 1 or $\partial_\tau$ in an associated equation within an adinkra, depending 
on whether the fermion or boson's associated node is above or below its super partner. 
For the bosons, we see the correct $\b{}_{{}_{\rm I}}$ assigning in 
Fig. \# 4. 

The numerical labels attached to each the nodes allow us to translate each diagram
into a matrix.  These are the absolute value of the color matrices.  We associate each bosonic nodal 
label with a row entry in a matrix and each fermionic nodal label with a column entry in 
a matrix.  Thus, for the bow tie decomposition, we obtain
\be\label{eq:BowtieB}
{\|\mathcal{B}}_{1L}\|= \left( \begin{array}{cc}
\beta_1^{-1} & 0\\ 
0 & \beta_1^{-1} \\ 
\end{array} \right)~~~,~~~~ 
 \|\mathcal{B}_{1R}\|= \left( \begin{array}{cc}
\beta_1& 0\\ 
0 & \beta_1 \\ 
\end{array} \right)~~~,~~~~ 
\ee
for the green color permutations and
\be
\|\mathcal{B}_{2L}\|= \left( \begin{array}{cc}
0 & \beta_2^{-1}\\ 
\beta_2^{-1} & 0 \\
\end{array} \right)  ~~~,~~~~ \\
 \|\mathcal{B}_{2R}\|= \left( \begin{array}{cc}
0 & \beta_2\\ 
\beta_2 & 0 \\
\end{array} \right)  ~~~, 
\ee
for the yellow line permutations. 

As mentioned earlier, we must take into account the line dashing as well,
in order to determine whether two adinkras are isomorphic. Therefore,
using Fig. \# 3, we find that the color matrices (instead of simply the 
absolute value of them) are:

\be\label{eq:BowtieB}
{\mathcal{B}}_{1L}= \left( \begin{array}{cc}
\beta_1^{-1} & 0\\ 
0 & \beta_1^{-1} \\ 
\end{array} \right)~~~,~~~~ 
 \mathcal{B}_{1R}= \left( \begin{array}{cc}
\beta_1& 0\\ 
0 & \beta_1 \\ 
\end{array} \right)~~~,~~~~ 
\ee
for the green color permutations and
\be
\mathcal{B}_{2L}= \left( \begin{array}{cc}
0 & -\beta_2^{-1}\\ 
\beta_2^{-1} & 0 \\
\end{array} \right)  ~~~,~~~~ \\
 \mathcal{B}_{2R} = \left( \begin{array}{cc}
0 & \beta_2\\ 
-\beta_2 & 0 \\
\end{array} \right)  ~~~, 
\ee
for the yellow line permutations. 

We associate 1 and $\partial_\tau$ respectively with  $\b{}_{{}_{\rm I}}$ 
and  $\b{}_{{}_{\rm I}}^{-1}$.  For the green lines in Fig.  \# \ref{f:diamond}, 
we find,
\be
{\rm D}{}_{{}_{\rm I}}
 \mathbf{\Phi}=\dot{\imath}\mathcal{B}{}_{{}_{\rm I}R} \mathbf{\Psi}
\label{Cmatrx1}
\ee
 where $\mathbf{\Phi}$ corresponds to the vector of bosons, and $\mathbf{\Psi}$ 
 corresponds to the vector of fermions.  In a similar manner, we have
 \be
{\rm D}{}_{{}_{\rm I}} \mathbf{\Psi}
=
\mathcal{B}{}_{{}_{\rm I}L}  \mathbf{\Phi}
\label{Cmatrx1a}
\ee
Therefore $\mathbf \Phi$ multiplied by  $ \mathcal{B}_{1R} \mathcal{B}_{2L}$ from 
the left is equivalent to the operation D${}_2$ D${}_1 \mathbf{\Phi}$, or similarly 
equivalent to following the green line and  then the yellow line from one boson to 
another. 

Next we can repeat all the steps above but now applied to the diamond adinkra
 (Fig \# 3).  
\begin{figure}[!htbp]
\centering
\includegraphics[width=0.8\columnwidth]{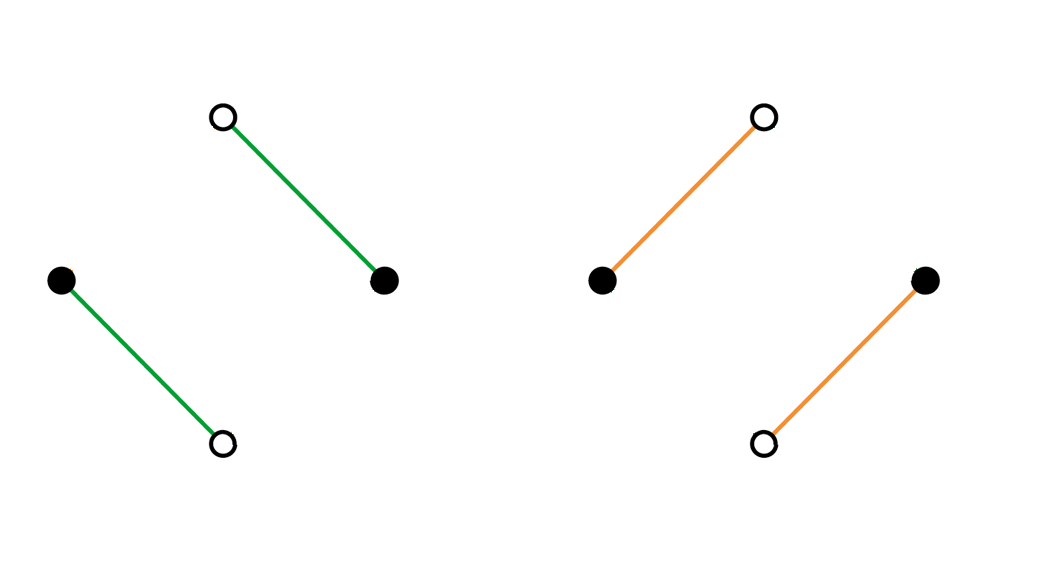}
\label{f:diamondN}
\caption{The monochromatic diamond edges.}
\end{figure} \newline
\noindent
It can easily be shown that the matrices are
\be\label{eq:DiamondB} {
{\mathcal{B}}_{1L}= \left( \begin{array}{cc}
\beta_1^{-1} & 0\\ 
0 & \beta_1 \\ 
\end{array} \right)~~~,~~~~ 
\mathcal{B}_{1R}= \left( \begin{array}{cc}
\beta_1 & 0\\ 
0 & \beta_1^{-1} \\ 
\end{array} \right)~~~,~~~~ 
}
\ee
for the green line permutations and
 \be
 \mathcal{B}_{2L}= \left( \begin{array}{cc}
0 & -\beta_2\\ 
\beta_2^{-1}  & 0 \\
\end{array} \right)   ~~~,~~~~\\
 \mathcal{B}_{2R}= \left( \begin{array}{cc}
0 & \beta_2\\ 
-\beta_2^{-1}  & 0 \\
\end{array} \right)   ~~~,
\label{Cmatrx2}
\ee
for the yellow line permutations.  Following the logic of the discussion given 
for the bowtie adinkra that led to (\ref{Cmatrx1}), an equation of the same
form can be obtained for the diamond adinkra.  The only difference is that
for the diamond the matrix ${\mathcal{B}}_{1L}$ is the one given in 
(\ref{eq:DiamondB}).
   
To make path tracing more explicit, let us first define the color dependent 
block matrix, which permute the super vector $\mathbf{\Phi}\oplus\mathbf{
\Psi}$ as defined below. \vskip0.1in
$\mathbf{Definition~1:}$\vskip0.1in
Let the matrix $\mathbf{C_j}$ be the color block matrix associated with the 
$j^{th}$ line color. We define
\vskip0.1in
\be
\mathbf{C_j}\equiv
\left( \begin{array}{cc}
0& \mathbf{\mathcal{B}_{jR}}\\ 
 \mathbf{\mathcal{B}_{jL}}&0\\ 
\end{array} \right)\ee
\vskip0.1in \noindent
where $\mathcal{B}_{jR}$ is the color matrix that permutes bosons to fermions 
and  $\mathcal{B}_{jL}$ is the color matrix the permutes fermions to bosons. \\

To create what we call an `isomorphism matrix', we can simply multiply all the 
color matrices from $\mathbf{\mathbf{C_N}}$ to $\mathbf{\mathbf{C_1}}$ from 
the left (which is equivalent to following a $N$-distinct color path with the 
specified color order from every node):

\be
\mathcal{B}=\mathbf{C_N} \mathbf{C_{N-1}} ... \mathbf{C_1}
\ee

Because every path can be covered by strictly looking at the $N$-distinct color 
paths from boson or fermion nodes alone, we claim that we can uniquely describe 
the adinkra up to line dashing by finding the eigenvalues in the matrix $\mathcal{
B}_{N(R/L)}...\mathcal{B}_{1R}$ where the first matrix is $\mathcal{B}_{NR}$ if 
there is an odd number of lines, and $\mathcal{B}_{NL}$ for an even number of 
lines.
  
If we were to inspect the eigenvalues of $\|\mathcal{B}_{1R}\mathcal{B}_{2L}\|$ 
in the diamond adinkra for example, we would find the eigenvalues are $\b_1 
\b_2$ and $\b_1^{-1}\b_2^{-1}$ while the bow tie adinkra has the eigenvalues 
$\pm \b_2\b_1^{-1}$, where the $\pm$ sign reflects the degenerate paths each 
boson takes in the bow tie. These two adinkras create distinct eigenvalues 
which agrees nicely with our statement above.

In general, finding the absolute value of the eivenvalues will determine adinkra shape,
while setting $\b_1,...,\b_N$ to 1, and taking the trace of $\mathcal{
B}_{N(R/L)}...\mathcal{B}_{1R}$ will determine the chromocharacters and hence the
dashing isometry. These two values is all one needs to determine whether two adinkras 
are isometric. For the adinkras in Fig. \# 4, we find the chromocharacters are 0, 
by setting $\b_1$ and $\b_2$ to 1.

Furthermore, we claim that relabeling nodes and re-defining the parity of nodes will not 
effect the absolute value of the eigenvalues (trivially the trace 
is basis independent). We can understand why by noting that permuting the adinkra
 nodes is equivalent to multiplying the equivalence matrix, $\mathcal{B}$ by
  \be
\mathcal{B}~~~\rightarrow~~~\mathcal{B}\left( \begin{array}{cc}
\mathbf{\mathcal{P}_1}& 0\\ 
0&\mathbf{\mathcal{P}_2}\\ 
\end{array} \right) \ee
\noindent
where $\mathcal{P}_i$ are permutation matrices. Because permutation matrices 
only move elements of the matrix around, and therefore do not affect eigenvalues,
$\mathcal{B}$'s eigenvalues remain the same.

The following theorem will explain why the algorithm always determines whether
adinkras are isomorphic.

$$ \mathbf{Theorem~1}$$

Two adinkras are isomorphic if and only if their associated eigenvalues of\newline 
$\|\mathcal{B}_{N(L/R)}\mathcal{B}_{[N-1](R/L)}...\mathcal{B}_{1R}\|$ (or equivalently 
$\|\mathcal{B}_{N(R/L)}...\mathcal{B}_{1L}\|$) are the same and their chromocharacters
are the same.
\vskip0.4in

$\mathbf{Proof:}$

$~~~$To prove the first part, we recall each eigenvalue carries information within the 
matrices $\|\mathcal{B}_{N(L/R)}\mathcal{B}_{[N-1](R/L)}...\mathcal{B
}_{1R}\|$ and $\|\mathcal{B}_{N(R/L)}...\mathcal{B}_{1L}\|$ that corresponds to the 
orientation upward or downward of a each colored line in an $N$-distinct color 
path. Because an adinkra has no more than one of each color at every node, a 
$N$-distinct color path that starts from strictly boson or strictly fermion nodes, 
will never share the same colored lines as another from the same type of node. 
Also, because every node has every color, every line would be included if we 
only record $N$-distinct color paths from bosons or fermions. 

If this was not the case, then there would exist a line that cannot be reached by an 
$N$-distinct color path from either type of node. Because every node has every 
color, however, every line color (besides the line itself) can connect to that 
`un-reachable' line from two directions. Therefore, there exists a fermion node 
and boson node that is less than length $N$ away such that, given an arbitrary 
order of path colors, it will reach the `un-reachable' line in a path of length less 
than $N$. Therefore, recording $N$-distinct color paths from the boson or fermion 
nodes will describe the movement of every colored line.
	
Now, we will prove that the $N$-distinct color paths are the same if and only if the
 adinkras are the same.
	
Trivially, if two adinkras are the same, then each $N$-distinct color path is the same.
 We are therefore left to prove the converse: if every $N$-distinct color path is the
  same, then the adinkras must be the same. If the adinkras were different, then there 
  would be a $N$-distinct color path between two nodes (of some arbitrary color 
  order) that is not the same between two adinkras. Because all the $N$-distinct 
  color paths include every line, this would imply at least one of the $N$-distinct 
  color paths with the color order $\|\mathbf{C_N} \mathbf{C_{N-1}} ... \mathbf{C_1}\|$
   is different, which is not possible.
   
   Lastly, to prove the second part, we recall that setting all the $\beta$ variables to 1
   and taking the trace of $\mathcal{B}_{N(L/R)}\mathcal{B}_{[N-1](R/L)}...\mathcal{B}_{1R}$
   or $\mathcal{B}_{N(R/L)}...\mathcal{B}_{1L}$ gives us the adinkra's associated
    chromocharacter, which determines line dashing isomorphisms.
    
  If two adinkras are the same shape, and have the same dashing isomorphisms, then
  trivially, they are isomorphic.
   
   \begin{flushright}$\Box$\end{flushright}
\vskip0.4in

To better understand the difference between the old and new algorithm,
 let us formally introduce the previous way of identifying isomorphism 
classes of adinkras.

\section{A Review of Past Adinkra Equivalence Algorithm}

$~~~$ Here we present the previous algorithm for determining the isomorphic class
 of a set of SUSY equations via an adinkra (This section was taken from \cite{Baobab}).

Firstly, we organize nodes according to distance from some arbitrary node $v$, and then
 organize each subset of nodes of a given distance by their lexicographically shortest path
  to $v$, as shown below.
\newline\newline
$\mathbf{Construction}$ $\mathbf{2}$: \newline
Given an adinkra, and the ordered set of colored lines $\{i_1,i_2,...i_{\mathcal{N}}\}$, let us
 choose an arbitrary node, $v$. We define the function $\lambda_v: V\rightarrow 
 [2^{\mathcal{N}-k}]$, as follows, where V is the set of adinkra vertices, and $k$ is the 
 number of non-hypercubic lines.

\begin{itemize}

\item $\lambda_v(v)=1$.

\item Order the nearest neighbors of $v$ from 2 to $\mathcal{N}+1$ based on the rank 
of the colored lines. In general, $\lambda_v(i_n v)=1+n, 1\le n\le N$.

\item Look at vertices a distance 2 away from $v$, and organize nodes equivalently,
 then the vertices a distance 3 away, and so on until all nodes are mapped to
  $[2^{\mathcal{N}-k}]$.

\end{itemize}

We will next inspect an adinkra with $h$ height assignments, and organize the $[2^k]$ set of 
orbits in an adinkra with a function $\tau$ which assigns the `$a^{th}$' orbit the number `$a$'.
 Next we define $\mu_g(h,a)$, which counts the number of nodes at a given height $h$ 
 in the orbit $a$:
\be
\mu_g(h,a)= |\{  v\in V: \text{hgt}(v)=h, \tau(v)=a \}|
\ee

where $\text{hgt}(v)$ is the relative height of the node from the bottom row. 
It can be shown that the certificate set,
\be
\text{cert}_G(x)=\{( \lambda_v(x),\text{hgt}(x),\tau(x)): x\in V\}
\ee
\noindent
can completely determine the isomorphism class of an adinkra \cite{DGW}. Despite the 
seeming simplicity, the authors did not demonstrate how to implement this algorithm 
except by tracing paths with your finger. With large adinkras (e.g. SUGRA $\mathcal{N}=32$
 adinkras), there is a great motivation to determine isomorphisms by computer instead of by hand. 
 How, one might ask, do we determine heights and orbits for a given adinkra computationally? 
 More importantly, how can we do this \emph{efficiently}? The previous paper left these
 glaring questions unanswered. 

This is the motivation behind efficiently distinguish isomorphism classes via color matrices.

\vskip0.7in
\section{On the Question of Efficiency}

$~~~$Given the previous algorithm, we run into a question regarding the relative efficiency
 of color matrix-based equivalence algorithms. 

Although our current algorithm only determines equivalence classes, we can easily use 
R- and L-matrices to determine cis- and trans adinkras, thus allowing the complete isomorphism
 class of the adinkra to be determined. Thus, we have an effective algorithm for determining
  isomorphism, but a quick look tells us we need a relatively enormous amount of memory to 
  determine isomorphisms compared to the previous algorithm.

Recall that we need a color matrix for every color with elements associated with every node. 
This gives $\cal N$ numbers for every node up to the value $\cal N$. Furthermore we need 
to denote $\pm$ signs for every R (L-matrices are transposes so they do not need to be recorded).
 This gives us $\mathcal{N}\text{log}_2(\mathcal{N})+\mathcal{N}/2$ bits to be recorded per node, or

\be
2^{\mathcal{N}-k}\[\mathcal{N}(\frac{1}{2} + \text{log}_2(\mathcal{N}))\]
\ee

bits total. In comparison, The previous algorithm needed 

\be
2^{\mathcal{N}-k}\[\mathcal{N}+\text{log}_2(\mathcal{N})\]
\ee

bits. Despite this disadvantage, the previous algorithm is impractical because there is no known
 computer-friendly way of determining orbits and levels \cite{Baobab}. More importantly, computer
  programs like Matlab are made for matrices, which reduces extra coding, and matlab code can 
  be added onto the GPU, allowing matrices to be created and computed with orders of magnitude 
  greater efficiency. It is because of these two distinct advantages that the algorithm has some 
  practical uses.

\section{Comparing Two Sets of Distinct Adinkras Using The 
Different Equivalence Algorithms}

$~\,~~$ Now we will apply this algorithm to two known adinkra sets, each of which 
have fundamentally distinct equations. In addition, we will show that each set of 
adinkras are distinct using the algorithm in \cite{DGW}, in order to demonstrate our 
algorithms superiority at determining adinkra equivalence with greater computational 
efficiency. 

\begin{figure}[!htbp]
   \centering
  \label{f:CE1}
    \includegraphics[width=0.8\columnwidth]{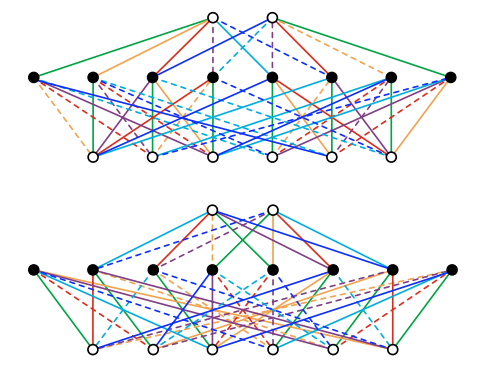}
     \caption{Our first set of adinkras being tested}
    \end{figure}    

In the adinkras, the following conventions are utilized: `1' is green, `2' is yellow, 
`3' is red, `4' is purple, `5' is light blue and `6' is dark blue. Because chromocharacters
 have been demonstrated to differentiate line dashing isomorphisms, we will simply look
 at the difference in adinkra shape with our algorithm. Here, both adinkras have the same
 dashing and hence have the same chromocharacters. For both sets of 
adinkras, we will list the color matrices $\|\mathbf{C_1}\|$ through $\|\mathbf{C_6}\|$, 
in Appendix A and Appendix B respectively, and compute the eigenvalues of 
$\|\mathcal{B}_{6L}...\mathcal{B}_{1R}\|$ and $\|\mathcal{B}_{6R} ...\mathcal{B}_{1L}\|$ 
below to demonstrate that the eigenvalues of two former matrices can adequately 
determine adinkra uniqueness, without having to resort to taking the eigenvalues 
of the entire $\mathcal{B}$ matrix.  

For the first adinkra in Figure \ref{f:CE1}, The eigenvalues for $\|\mathcal{B}_{6L
}...\mathcal{B}_{1R}\|$ are 
\be
\pm 1 \, (\text{2x degenerate}) ~~,~~ 
 \pm \b_2 \b_4 \b_6 \b_1^{-1}\b_3^{-1} \b_5^{-1}  \, (\text{2x degenerate}) ~~~,
\ee
and the eigenvalues for $\|\mathcal{B}_{6R}...\mathcal{B}_{1L}\|$ are
\be
\pm \beta_1\b_2^{-1}  \, (\text{2x degenerate}) 
~~,~~  \pm \beta_3 \beta_5 \beta_4^{-1}\beta_6^{-1}   \, (\text{2x degenerate})
~~~.
\ee\indent
For the second adinkra, the eigenvalues for $\|\mathcal{B}_{6L}...\mathcal{B}_{
1R}\|$ are 
\be
\pm\beta_2 \beta_4\beta_6\beta_1^{-1}\beta_3^{-1}\beta_5^{-1} 
(\text{3x degenerate}) ~~,~~
\pm\beta_1\beta_3\beta_5\beta_2^{-1} \beta_4^{-1}\beta_6^{-1} ~~,
\ee
and the eigenvalues for $\|\mathcal{B}_{6R}...\mathcal{B}_{1L}\|$ in the second 
adinkra  are
\be  \eqalign{
&\pm\beta_1\beta_3\beta_5\beta_2^{-1} \beta_4^{-1}\beta_6^{-1}  ~~,~~ 
\pm\beta_2\beta_3\beta_5\beta_1^{-1} \beta_4^{-1}\beta_6^{-1} ~~~,  \cr 
&\pm\beta_1\beta_4\beta_5\beta_2^{-1} \beta_3^{-1}\beta_6^{-1} ~~,~~ 
\pm\beta_1\beta_3\beta_6\beta_2^{-1} \beta_4^{-1}\beta_5^{-1} ~~~.
 }  \ee
We note that the sets of eigenvalues are distinct from the previous adinkra.

The previous algorithm used the following set
\be
cert_G(v)=\{(\lambda_v(x),hgt(x), \tau(x)):x\in V\}
\ee
to determine equivalence, where $\lambda_v(x)$ organizes the nodes to 
determine if two adinkras are the same ignoring dashing or automorhpisms. 
$hgt(x)$ determines the absolute height of the node (e.g. 2 rows from the 
bottom row). $\tau(x)$ is used to determine the orbits of an adinkra.

\begin{figure}[!htbp]
   \centering
  \label{f:CE2}
    \includegraphics[width=0.8\columnwidth]{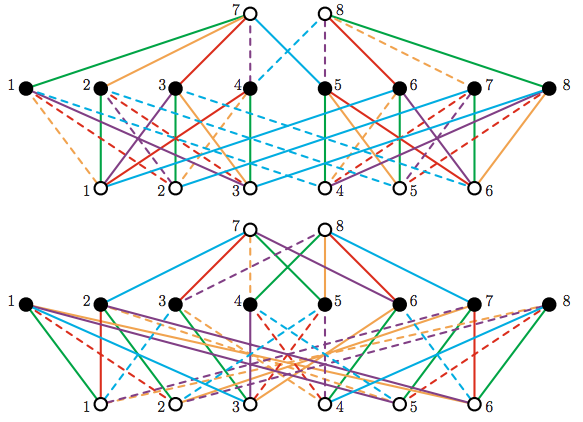}
     \caption{The second set of adinkras being tested}
\end{figure}  

Our next adinkra pair in Fig. \# \ref{f:CE2} is the same as the previous 
except without the dark blue lines, hence the color matrices $C_J$ 
are the same except there is no $\mathbf{C_6}$.

For the first adinkra, the eigenvalues of $\|\mathcal{B}_{5R}...\mathcal{
B}_{1R}\|$ are 

\begin{center}
$\beta_1^{1/3}\beta_ 3^{1/3}\beta_5^{1/3}\beta_2^{-1}\beta_4^{-1/3}, 
\  $\vskip0.1in
$-(-1)^{1/3}\beta_1^{1/3}\beta_ 3^{1/3}\beta_5^{1/3}\beta_2^{-1}\beta_4^{-1/3}, \  $\vskip0.1in
$(-1)^{2/3}\beta_1^{1/3}\beta_ 3^{1/3}\beta_5^{1/3}\beta_2^{-1}\beta_4^{-1/3}, \ $\vskip0.1in
$\beta_1^{1/5}\beta_ 3^{3/5}\beta_5^{3/5}\beta_2^{-1/5}\beta_4^{-3/5}, \  $\vskip0.1in
$-(-1)^{1/5}\beta_1^{1/5}\beta_ 3^{3/5}\beta_5^{3/5}\beta_2^{-1/5}\beta_4^{-3/5}, \ $\vskip0.1in
$(-1)^{2/5}\beta_1^{1/5}\beta_ 3^{3/5}\beta_5^{3/5}\beta_2^{-1/5}\beta_4^{-3/5}, \ $\vskip0.1in
$-(-1)^{3/5}\beta_1^{1/5}\beta_ 3^{3/5}\beta_5^{3/5}\beta_2^{-1/5}\beta_4^{-3/5}, \ $\vskip0.1in
\end{center}
\be
(-1)^{4/5}\beta_1^{1/5}\beta_ 3^{3/5}\beta_5^{3/5}\beta_2^{-1/5}\beta_4^{-3/5}
\ee

The eigenvalues of $\|\mathcal{B}_{5L}...\mathcal{B}_{1L}\|$ are 
\begin{center}
$\beta_2^{1/5}\beta_4^{3/5}\beta_1^{-3/5}\beta_3^{-1/5}\beta_5^{-3/5}, \ $\vskip0.1in
$-(-1)^{1/5}\beta_2^{1/5}\beta_4^{3/5}\beta_1^{-3/5}\beta_3^{-1/5}\beta_5^{-3/5}, \ $ \vskip0.1in
$(-1)^{2/5}\beta_2^{1/5}\beta_4^{3/5}\beta_1^{-3/5}\beta_3^{-1/5}\beta_5^{-3/5}, \ $\vskip0.1in
$-(-1)^{3/5}\beta_2^{1/5}\beta_4^{3/5}\beta_1^{-3/5}\beta_3^{-1/5}\beta_5^{-3/5}, \ $\vskip0.1in
$(-1)^{4/5}\beta_2^{1/5}\beta_4^{3/5}\beta_1^{-3/5}\beta_3^{-1/5}\beta_5^{-3/5}, \ $\vskip0.1in            
$\beta_2\beta_4^{1/3}\beta_1^{-1/3}\beta_3^{-1}\beta_5^{-1/3}, \ $ \vskip0.1in
$-(-1)^{1/3}\beta_2\beta_4^{1/3}\beta_1^{-1/3}\beta_3^{-1}\beta_5^{-1/3}, \ $\vskip0.1in
\end{center}
\be
(-1)^{2/3}\beta_2\beta_4^{1/3}\beta_1^{-1/3}\beta_3^{-1}\beta_5^{-1/3}
\ee

In comparison, for the second adinkra, the eigenvalues of $\|\mathcal{B}_{5R}...\mathcal{B}_{1R}\|$ are 
\begin{center}
$\beta_1^{1/5}\beta_3\beta_5^{1/5}\beta_2^{-1/5}\beta_4^{-1}, \ $\vskip0.1in
$-(-1)^{1/5}\beta_1^{1/5}\beta_3\beta_5^{1/5}\beta_2^{-1/5}\beta_4^{-1}, \ $\vskip0.1in
$(-1)^{2/5}\beta_1^{1/5}\beta_3\beta_5^{1/5}\beta_2^{-1/5}\beta_4^{-1}, \ $\vskip0.1in
$-(-1)^{3/5}\beta_1^{1/5}\beta_3\beta_5^{1/5}\beta_2^{-1/5}\beta_4^{-1}, \ $\vskip0.1in
$(-1)^{4/5}\beta_1^{1/5}\beta_3\beta_5^{1/5}\beta_2^{-1/5}\beta_4^{-1}, \ $\vskip0.1in
$-\beta_1\beta_3\beta_5\beta_2^{-1}\beta_4^{-1}, \ $\vskip0.1in
\end{center}
\be\pm\beta_1\beta_5\beta_2^{-1} \ee

The eigenvalues for  $\|\mathcal{B}_{5L}...\mathcal{B}_{1L}\|$ are 
\begin{center}
$\beta_2^{1/5}\beta_4\beta_1^{-1/5}\beta_3^{-1}\beta_5^{-1/5}, \ $\vskip0.1in
$-(-1)^{1/5}\beta_2^{1/5}\beta_4\beta_1^{-1/5}\beta_3^{-1}\beta_5^{-1/5}, \ $\vskip0.1in
$(-1)^{2/5}\beta_2^{1/5}\beta_4\beta_1^{-1/5}\beta_3^{-1}\beta_5^{-1/5}, \  $\vskip0.1in
$-(-1)^{3/5}\beta_2^{1/5}\beta_4\beta_1^{-1/5}\beta_3^{-1}\beta_5^{-1/5}, \  $\vskip0.1in
$(-1)^{4/5}\beta_2^{1/5}\beta_4\beta_1^{-1/5}\beta_3^{-1}\beta_5^{-1/5}, \ $\vskip0.1in
$-\beta_2\beta_4\beta_1^{-1}\beta_3^{-1}\beta_5^{-1},\ $
\end{center}
\be\beta_2\beta_4\beta_1^{-1}\beta_3^{-1}\beta_5^{-1} (\text{3x degenerate}) 
\ee

Using the old algorithm, we find that although $\lambda(x)$ is the same, we find that

$
~~~~~~~~~~~~\mu_G(3,1) = 1 ~~~~~~~~~~~~~~~~~~~~~~~~~~~~~~~ \mu_H(3,1)=2 \newline
~~~~~~~~~~~~~~~~~\mu_G(3,2) = 1 ~~~~~~~~~~~~~~~~~~~~~~~~~~~~~~~~ \mu_H(3,2)=0 \newline
~~~~~~~~~~~~~~~~~\mu_G(2,1)=\mu_G(2,2) = 4 ~~~~~~~~~~~~~~~~~~ \mu_H(2,1)=\mu_H(2,2)=4 \newline
~~~~~~~~~~~~~~~~~\mu_G(1,1) = 3 ~~~~~~~~~~~~~~~~~~~~~~~~~~~~~~~~ \mu_H(1,1)=2 \newline
~~~~~~~~~~~~~~~~~\mu_G(1,2) =3 ~~~~~~~~~~~~~~~~~~~~~~~~~~~~~~~~\mu_H(1,2)=4
$ \newline \noindent
and clearly the the automorphisms are not the same. This, again, took significantly 
longer to determine both because nodes had to be organized, and needs height 
determination to generally determine the in-equivalence of two adinkras, which 
is again not necessary for our current algorithm

\section{Conclusion}

$~\,~~$ In this paper, we presented a new algorithm to determine whether two adinkras
are isomorphic, using the trace of of $\mathcal{B}_{N(R/L)}...\mathcal{B}_{1R}$, setting all
 variables to 1, and by using the eigenvalues of the absolute value of the above matrix.
These eigenvalues correspond to $N$-length unique color paths from bosons 
(if $\mathcal{B}_{N R}$), or fermions (if $\mathcal{B}_{N L}$). It is clear 
that not only is the algorithm efficient, but can disregard the order of the node labels. 
Furthermore, despite the comparative inefficiency of this algorithm compared to \cite{DGW}
the algorithm is much more computer friendly, and can be computed with parallel processors
saving even more time in the process.
$$~~$$

${~~~}$ \newline
${~~~~~}$``{\it {A good picture is equivalent to a good deed.
}}"${~~~}$ \newline
\newline $~~~~~~~$ -- Vincent Van Gogh
\newline ${~~~}$ \newline
\vskip0.5in
\noindent
{\Large\bf Acknowledgments}

This research was supported in part by the endowment of the John S.~Toll 
Professorship, the University of Maryland Center for String \& Particle Theory, 
National Science Foundation Grant PHY-09-68854.  Additionally KB 
acknowledges participation in  2012 SSTPRS (Student Summer Theoretical 
Physics Research Session).  Adinkras were drawn with \emph{Adinkramat} 
$\copyright$ 2008 by G. Landweber.

\newpage
\section{Appendix A}

The color matrices, $C_J$, for the first adinkra in Fig. \# \ref{f:CE1} are listed below. For all the adinkras in Appendix B and C, I used the convention: ``1" is green, ``2" is yellow, ``3" is red, ``4" is purple, ``5" is light blue and ``6" is dark blue. The color matrices for the first adinkra in Fig. \# \ref{f:CE2} are the same, except without the final matrix, $\mathbf{C_6}$.
\vskip0.1in

$$
\vCent
 {\setlength{\unitlength}{1mm}
  \begin{picture}(-20,0)
   \put(-77,-88){\includegraphics[width=1\columnwidth]{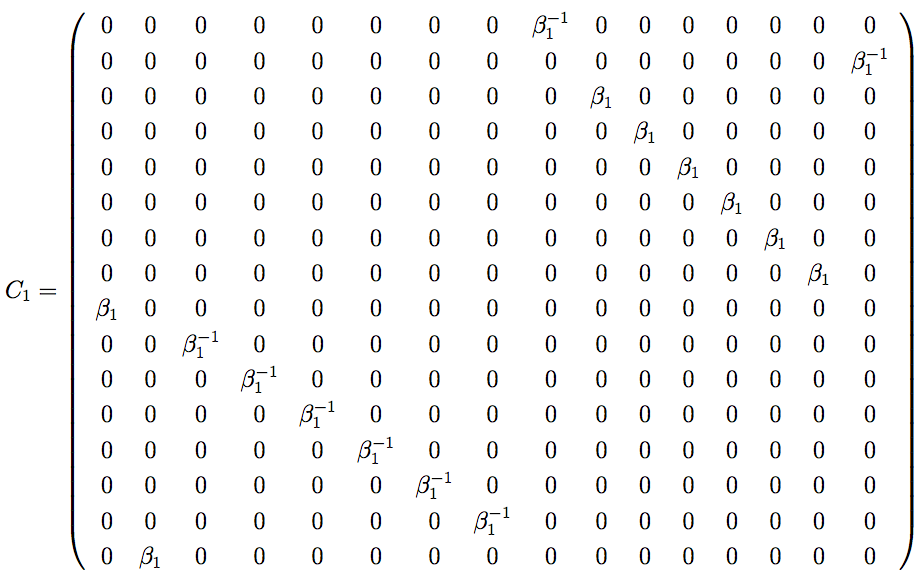}}
  \end{picture}}.
$$ 
\vskip3.2in
$$
\vCent
 {\setlength{\unitlength}{1mm}
  \begin{picture}(-20,0)
   \put(-77,-88){\includegraphics[width=1\columnwidth]{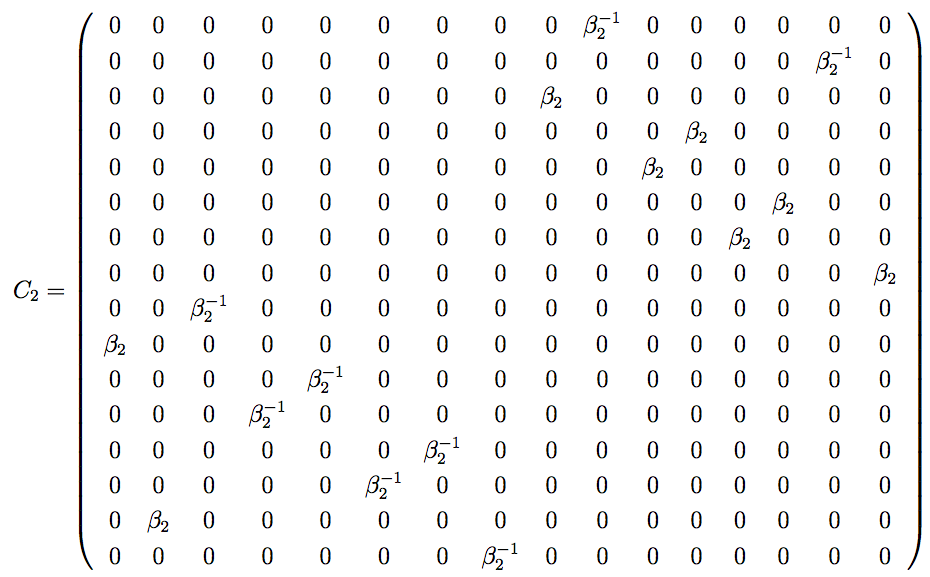}}
  \end{picture}}.
$$ 
\vskip5.2in
$$
\vCent
 {\setlength{\unitlength}{1mm}
  \begin{picture}(-20,0)
   \put(-77,-88){\includegraphics[width=1\columnwidth]{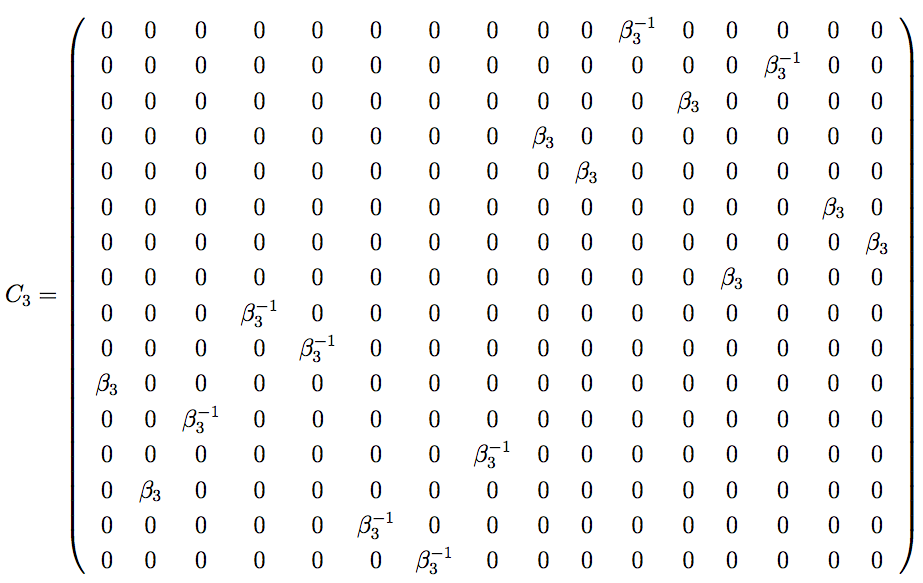}}
  \end{picture}}.
$$ 
\vskip3.2in
$$
\vCent
 {\setlength{\unitlength}{1mm}
  \begin{picture}(-20,0)
   \put(-77,-88){\includegraphics[width=1\columnwidth]{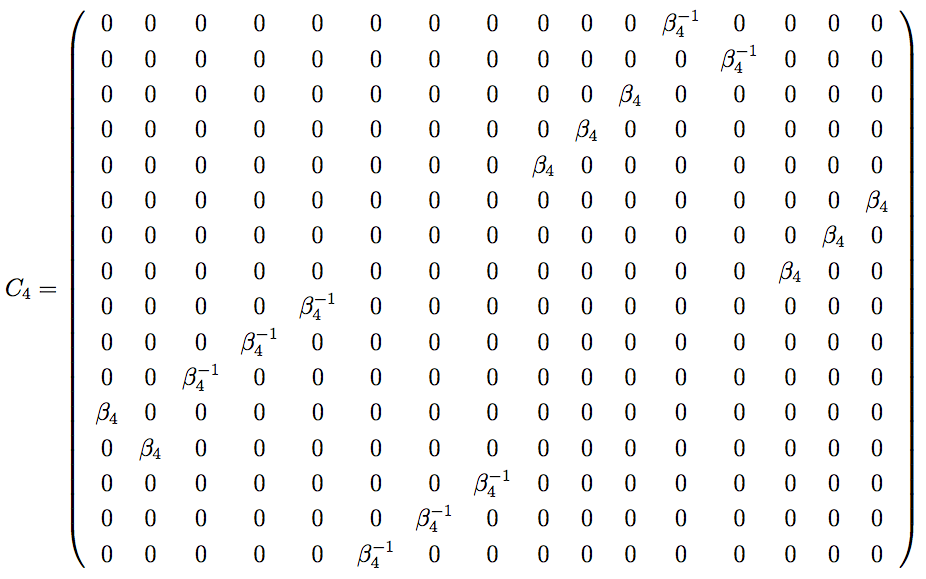}}
  \end{picture}}.
$$ 
\vskip5.2in
$$
\vCent
 {\setlength{\unitlength}{1mm}
  \begin{picture}(-20,0)
   \put(-77,-88){\includegraphics[width=1\columnwidth]{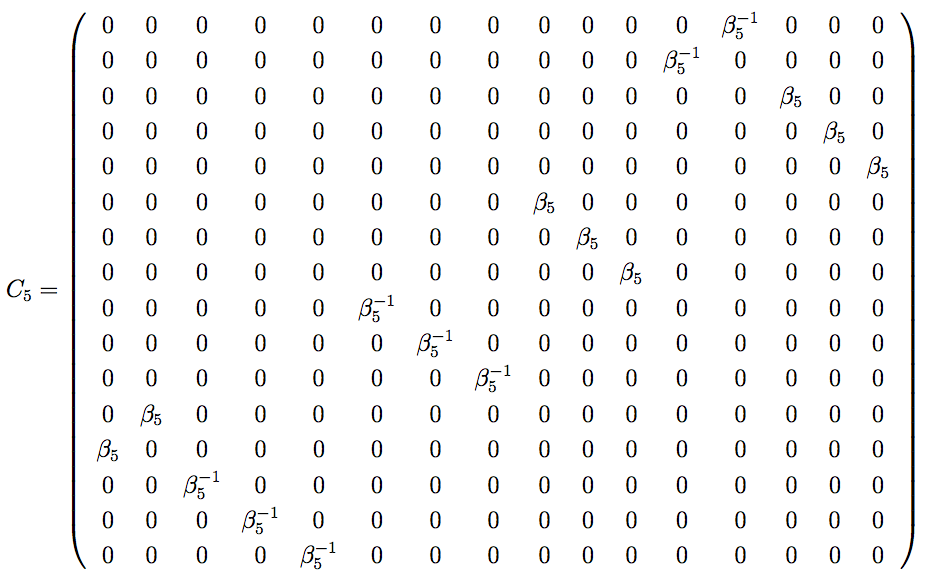}}
  \end{picture}}.
$$ 
\vskip3.2in
$$
\vCent
 {\setlength{\unitlength}{1mm}
  \begin{picture}(-20,0)
   \put(-77,-88){\includegraphics[width=1\columnwidth]{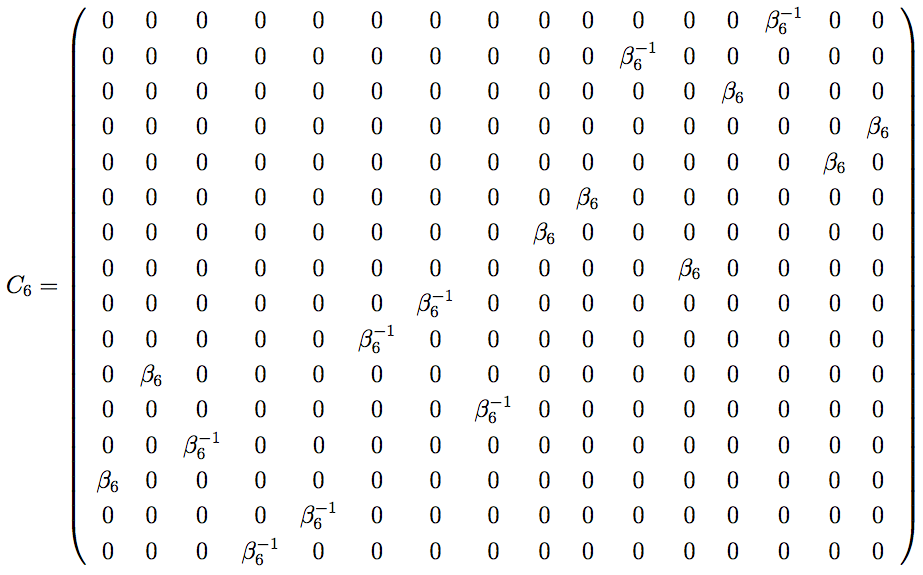}}
  \end{picture}}.
$$ 
\vskip3.2in

\newpage
\section{Appendix B}

The color matrices, $C_J$, for the second adinkra in Fig. \# \ref{f:CE1} are listed below. For all the adinkras in Appendix B and C, I used the convention: ``1" is green, ``2" is yellow, ``3" is red, ``4" is purple, ``5" is light blue and ``6" is dark blue. The color matrices for the second adinkra in Fig. \# \ref{f:CE2} are the same, except without the final matrix, $\mathbf{C_6}$.

\vskip1.5in
$$
\vCent
 {\setlength{\unitlength}{1mm}
  \begin{picture}(-20,0)
   \put(-77,-48){\includegraphics[width=1\columnwidth]{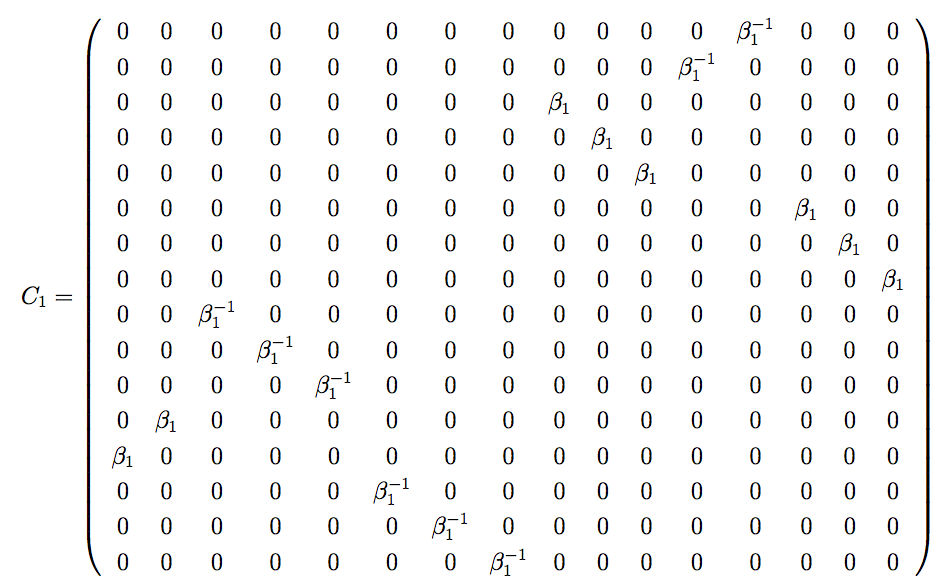}}
  \end{picture}}.
$$ 
\vskip3.2in

$$
\vCent
 {\setlength{\unitlength}{1mm}
  \begin{picture}(-20,0)
   \put(-77,-48){\includegraphics[width=1\columnwidth]{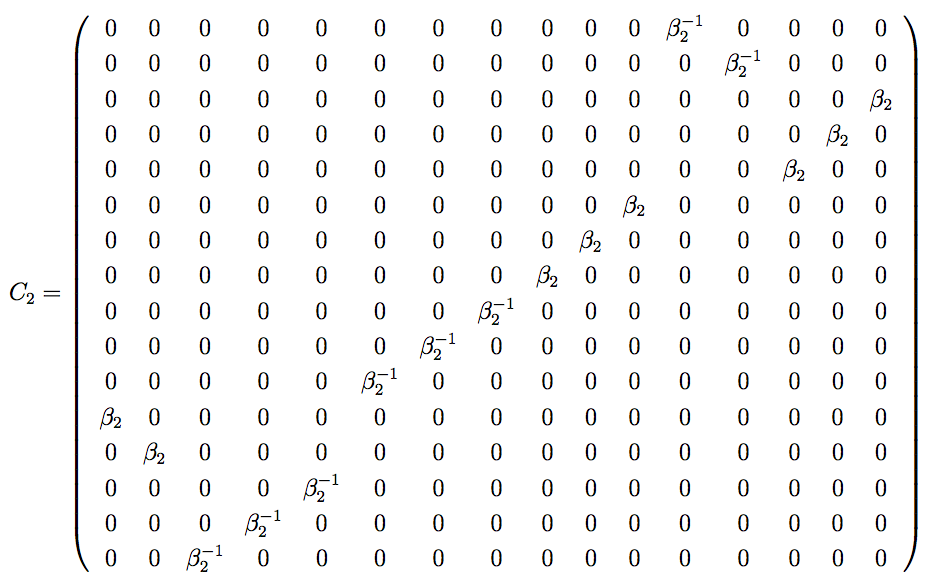}}
  \end{picture}}.
$$ 
\vskip3.2in
$$
\vCent
 {\setlength{\unitlength}{1mm}
  \begin{picture}(-20,0)
   \put(-77,-88){\includegraphics[width=1\columnwidth]{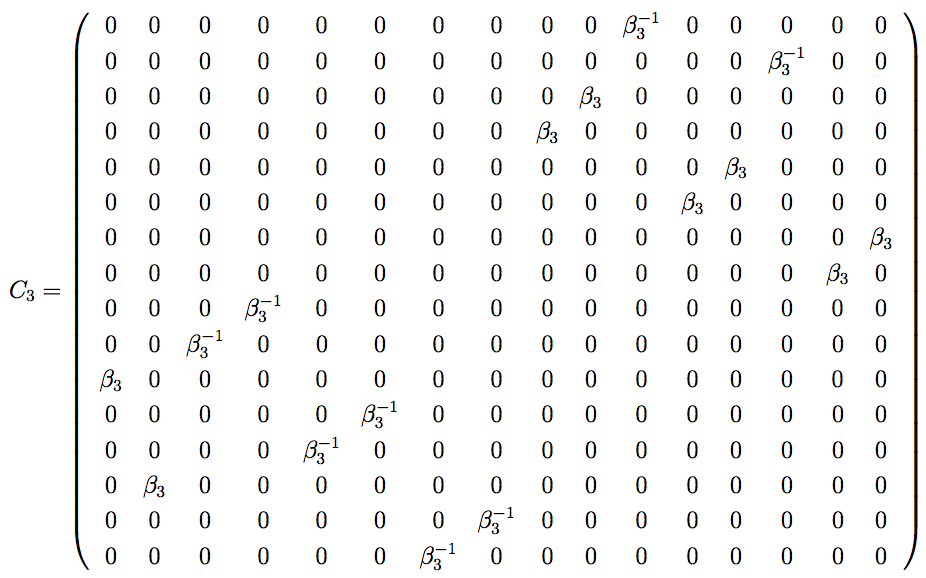}}
  \end{picture}}.
$$ 
\vskip3.2in

$$
\vCent
 {\setlength{\unitlength}{1mm}
  \begin{picture}(-20,0)
   \put(-77,-88){\includegraphics[width=1\columnwidth]{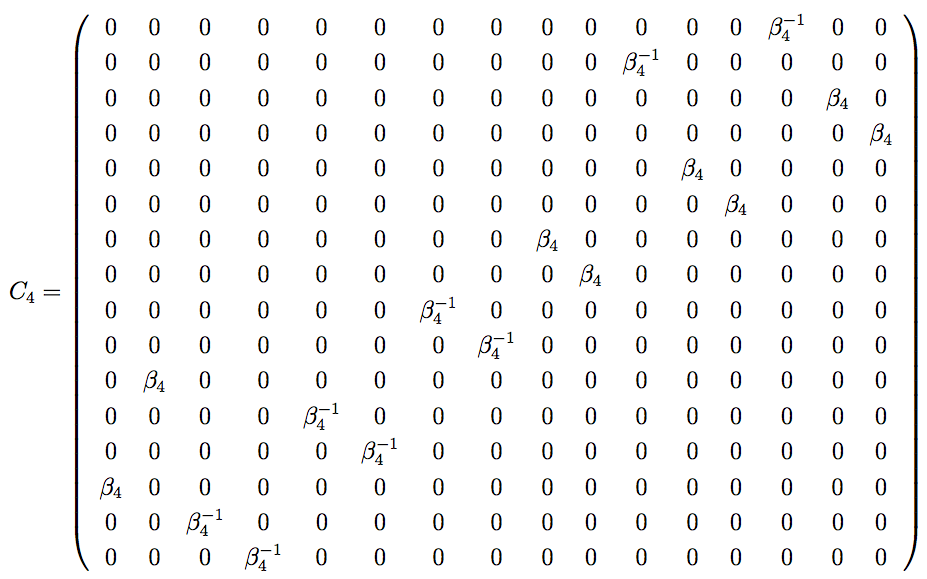}}
  \end{picture}}.
$$ 
\vskip5.2in
$$
\vCent
 {\setlength{\unitlength}{1mm}
  \begin{picture}(-20,0)
   \put(-77,-88){\includegraphics[width=1\columnwidth]{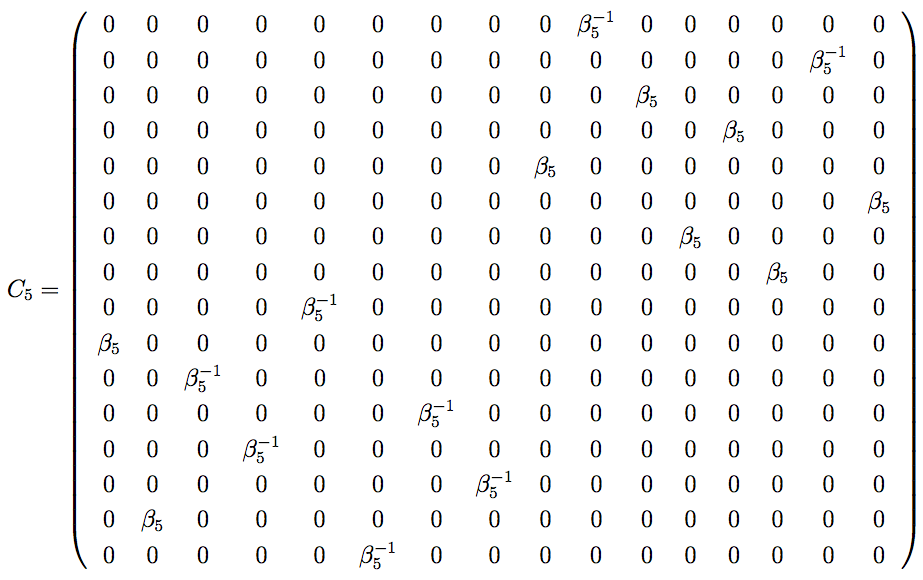}}
  \end{picture}}.
$$ 
\vskip3.2in

$$
\vCent
 {\setlength{\unitlength}{1mm}
  \begin{picture}(-20,0)
   \put(-77,-88){\includegraphics[width=1\columnwidth]{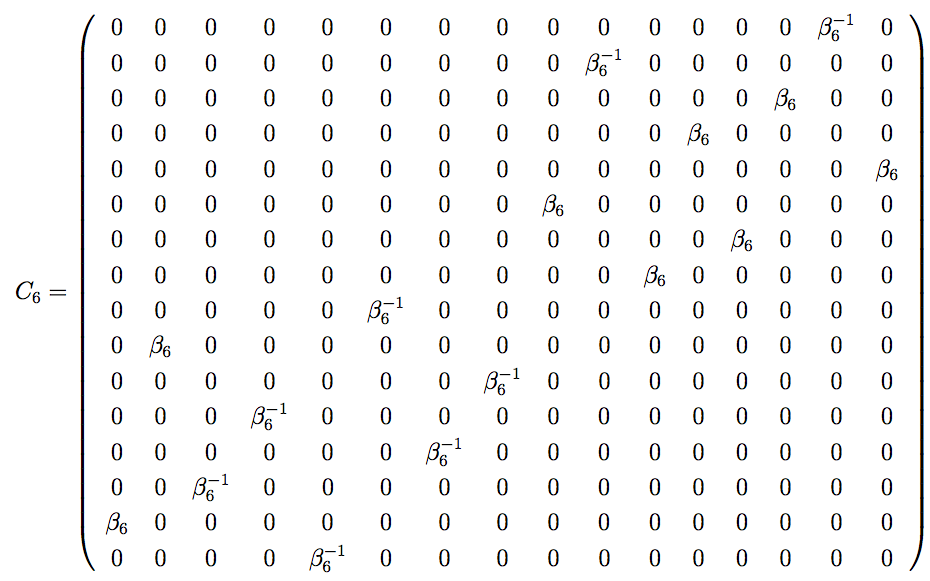}}
  \end{picture}}.
$$ 
\vskip3.2in


\end{document}